\RequirePackage{fix-cm} 
\documentclass[a4paper, twoside, reqno, dvips, 12pt]{amsart}
\usepackage{fixltx2e}   

\usepackage[latin1]{inputenc}
\usepackage[T1]{fontenc}

\usepackage{eucal}
\usepackage{esint}
\usepackage{dsfont}
\usepackage{xspace}
\usepackage{amsgen}
\usepackage{amsthm}
\usepackage{amssymb}
\usepackage{amsmath}
\usepackage{upgreek}
\usepackage{amsfonts}
\usepackage{MnSymbol}
\usepackage{mathrsfs}
\usepackage{mathtools}
\usepackage[nice]{nicefrac}

\usepackage{a4wide}

\usepackage[table]{xcolor}

\definecolor{oneblue}{rgb}{0.0, 0.0, 0.85}
\definecolor{darkgrey}{rgb}{0.273, 0.281, 0.30}
\definecolor{Lightgray}{rgb}{0.89, 0.89, 0.89}
\definecolor{Lightblue}{RGB}{214, 214, 214}
\definecolor{bckg}{RGB}{20.8, 20.8, 20.8} 

\usepackage{psfrag}
\usepackage{graphicx}
\usepackage{subfigure}
\usepackage{morefloats}
\usepackage{indentfirst}

\usepackage{acronym}
\usepackage{microtype}
\usepackage[labelfont={bf,sf},%
            labelsep=period,%
            justification=raggedright]{caption}

\usepackage[usenames, dvipsnames, pdf]{pstricks}
\usepackage{epsfig}
\usepackage{pst-grad} 
\usepackage{pst-plot} 

\usepackage[colorlinks,
            urlcolor=oneblue,
            linkcolor=oneblue,
            citecolor=oneblue,
            bookmarksopen=false,
            pdfpagemode=UseNone,
            pagebackref]{hyperref}

\usepackage[explicit]{titlesec}

\titleformat{\section}
  {\color{darkgrey}\Large\sffamily\bfseries}
  {}
  {0em}
  {\colorbox{bckg!5}{\parbox{\dimexpr\linewidth-2\fboxsep\relax}{\centering\thesection. #1}}}
  []

\titleformat{name=\section,numberless}
  {\color{darkgrey}\Large\sffamily\bfseries}
  {}
  {0em}
  {\colorbox{bckg!10}{\parbox{\dimexpr\linewidth-2\fboxsep\relax}{\centering#1}}}
  []

\titleformat{\subsection}
  {\color{darkgrey}\large\sffamily\bfseries}
  {}
  {0em}
  {\colorbox{bckg!5}{\parbox{\dimexpr\linewidth-2\fboxsep\relax}{\centering\thesubsection. #1}}}
  []

\titleformat{name=\subsection,numberless}
  {\color{darkgrey}\Large\sffamily\bfseries}
  {}
  {0em}
  {\colorbox{bckg!10}{\parbox{\dimexpr\linewidth-2\fboxsep\relax}{\centering#1}}}
  []

\titleformat{\subsubsection}
  {\sffamily\normalsize\bfseries}
  {\thesubsubsection}
  {0.5em}
  {#1}
  []

\titleformat{\paragraph}[runin]
  {\sffamily\small\bfseries}
  {}
  {0em}
  {#1}

\titlespacing*{\section}{1.0em}{1.0em}{0.8em}[0em]
\titlespacing*{\subsection}{1.0em}{1.0em}{0.8em}[0em]
\titlespacing*{\subsubsection}{1.0em}{0.7em}{0.6em}[0em]

\usepackage{titletoc}

\contentsmargin{0.5em}
\newlength{\tocsep} 
\titlecontents{section}[\tocsep]
  {\addvspace{10pt}\bfseries\sffamily}
  {\contentslabel[\thecontentslabel]{\tocsep}}
  {}
  {\ \titlerule*[0.75pc]{.}\ \thecontentspage}
  []
\titlecontents{subsection}[\tocsep]
  {\addvspace{8pt}\sffamily}
  {\contentslabel[\thecontentslabel]{\tocsep}}
  {}
  {\ \titlerule*[0.5pc]{.}\ \thecontentspage}
  []
\titlecontents*{subsubsection}[\tocsep]
  {\footnotesize\sffamily}
  {}
  {}
  {\ \titlerule*[0.35pc]{.}\ \thecontentspage}
  [\ \textbullet\ ]  

\usepackage{fancyhdr}
\usepackage{lastpage}

\newcommand*\Title{Adaptive shallow water modeling}
\newcommand*\Authors{D.~Dutykh, D.~Clamond \& D.~Mitsotakis}
\newcommand*{\plogo}{{\texttt{arXiv.org} / \textsc{hal}}} 

\numberwithin{equation}{section}

\newtheorem{remark}{Remark}

\newcommand{\R}{\mathbb{R}}

\newcommand{\ud}{\mathrm{d}}
\newcommand{\ui}{\mathrm{i}}
\newcommand{\ue}{\mathrm{e}}
\newcommand{\F}{\mathcal{F}}

\newcommand{\eps}{\varepsilon}
\renewcommand{\O}{\mathcal{O}}

\renewcommand{\S}{\mathcal{S}}
\renewcommand{\alpha}{\upalpha}

\renewcommand{\gamma}{\boldsymbol{\upgamma}}

\newcommand{\ie}{\emph{i.e.}~}
\newcommand{\eg}{\emph{e.g.}~}
\newcommand{\etal}{\emph{et al.}~}

\newcommand{\sech}{\mathrm{sech}}

\newcommand{\half}{{\textstyle{1\over2}}}
\newcommand{\third}{{\textstyle{1\over3}}}
\newcommand{\sixth}{{\textstyle{1\over6}}}
\newcommand{\twothird}{{\textstyle{2\over3}}}

\usepackage{acronym}
\acrodef{SGN}[SGN]{Serre--Green--Naghdi}
\acrodef{eSGN}[eSGN]{extended Serre--Green--Naghdi}

\begin{document}

\title[\Title]{Adaptive modeling of shallow fully nonlinear gravity waves}

\author[D. Dutykh]{Denys Dutykh$^*$}
\address{LAMA, UMR 5127 CNRS, Universit\'e Savoie Mont Blanc, Campus Scientifique, 73376 Le Bourget-du-Lac Cedex, France}
\email{Denys.Dutykh@univ-savoie.fr}
\urladdr{http://www.lama.univ-savoie.fr/~dutykh/}
\thanks{$^*$ Corresponding author}

\author[D. Clamond]{Didier Clamond}
\address{Laboratoire J.-A. Dieudonn\'e, Universit\'e de Nice -- Sophia Antipolis, Parc Valrose, 06108 Nice cedex 2, France}
\email{diderc@unice.fr}
\urladdr{http://math.unice.fr/~didierc/}

\author[D.~Mitsotakis]{Dimitrios Mitsotakis}
\address{Victoria University of Wellington, School of Mathematics, Statistics and Operations Research, PO Box 600, Wellington 6140, New Zealand}
\email{dmitsot@gmail.com}
\urladdr{http://dmitsot.googlepages.com/}


\begin{titlepage}
\setcounter{page}{1}
\thispagestyle{empty} 
\noindent
{\Large Denys \textsc{Dutykh}}\\
{\it\textcolor{gray}{CNRS, Universit\'e Savoie Mont Blanc, France}}\\[0.02\textheight]
{\Large Didier \textsc{Clamond}}\\
{\it\textcolor{gray}{Universit\'e de Nice -- Sophia Antipolis, France}}\\[0.02\textheight]
{\Large Dimitrios \textsc{Mitsotakis}}\\
{\it\textcolor{gray}{Victoria University of Wellington, New Zealand}}\\[0.16\textheight]

\colorbox{Lightblue}{
  \parbox[t]{1.0\textwidth}{
    \centering\huge\sc
    \vspace*{0.7cm}
    
    Adaptive modeling of shallow fully nonlinear gravity waves

    \vspace*{0.7cm}
  }
}

\vfill 

\raggedleft     
{\large \plogo} 
\end{titlepage}


\begin{abstract}

This paper presents an extended version of the celebrated \acf{SGN} system. This extension is based on the well-known Bona--Smith--Nwogu trick which aims to improve the linear dispersion properties. We show that in the fully nonlinear setting it results in modifying the vertical acceleration. Even if this technique is well-known, the effect of this modification on the nonlinear properties of the model is not clear. The first goal of this study is to shed some light on the properties of solitary waves, as the most important class of nonlinear permanent solutions. Then, we propose a simple adaptive strategy to choose the \emph{optimal value} of the free parameter at every instance of time. This strategy is validated by comparing the model prediction with the reference solutions of the full Euler equations and its classical counterpart. Numerical simulations show that the new adaptive model provides a much better accuracy for the same computational complexity.

\bigskip
\noindent \textbf{\keywordsname:} Shallow water waves; Serre equations; Green--Naghdi model; dispersion relation; adaptive model; solitary waves \\


\end{abstract}


\newpage
\tableofcontents
\thispagestyle{empty}


\newpage
\section{Introduction}

The water wave theory has always been developed through the derivation and analysis of various approximate models \cite{Craik2004}. Nowadays the researchers, motivated by practical or theoretical needs, continue actively the quest for more accurate simplified models. In the present study our starting point is a celebrated set of equations which was derived for the first time by F.~\textsc{Serre} \cite{Serre1953} in 1953, even if a deeper literature search shows that a steady version of Serre's equations were already present in works of Lord \textsc{Rayleigh} (1876) \cite{LordRayleigh1876}. Then, this system was rediscovered independently by \textsc{Su} \& \textsc{Gardner} (1969) \cite{SG1969}, and again by \textsc{Green}, \textsc{Laws} \& \textsc{Naghdi} (1974) \cite{Green1974}. In the Soviet literature this model was known as the Zheleznyak--Pelinovsky model \cite{Zheleznyak1985}. The derivation of these equations from variational principles was given in \cite{Miles1985, Kim2001, Clamond2009}. This list of references is far from being exhaustive. In the rest of the manuscript we will refer to this set of equations as the \acf{SGN} system.

The \acs{SGN} equations are fully nonlinear but only weakly dispersive \cite{Lannes2009}. Consequently, one could think how to improve the dispersive characteristics of the model \cite{Dias2010}. Fortunately, some technology has already been developed for the Boussinesq-type equations. The technique of introducing the free parameters into long wave models was pioneered by \textsc{Bona} \& \textsc{Smith} (1976) \cite{BS} and later independently by \textsc{Nwogu} (1993) \cite{Nwogu1993}. The idea mainly consists in using the horizontal velocity variable defined at the arbitrary depth along with lower order asymptotic relations to alter higher order terms. This technique was synthesized by \textsc{Bona} \etal (2002) \cite{BCS}. There is also another approach based on Pad\'e-type approximations due to \textsc{Madsen} and his collaborators \cite{Madsen2002, Madsen03}. All these methods have been succesfully employed to derive various extended Boussinesq-type equations \cite{Beji1996, Liu2005, Kim2009}.

Once the model equations have been proposed, one has to propose also efficient way to solve them numerically. Currently, there is an important research activity towards the numerical solution of various Boussinesq-type dispersive wave models. Only recently various finite volume \cite{ChazelLannes2010, Kazolea2013, Dutykh2011a}, finite element (FEM/continuous Galerkin) \cite{DMS1, Mitsotakis2014}, pseudo-spectral \cite{Dutykh2011a}, discontinuous Galerkin \cite{Eskilsson2006} and residual distribution \cite{Ricchiuto2014} schemes have been proposed.

However, for practical simulations the free parameters have to be assigned with some values. It is obvious also that different choices may lead to models with completely different properties. Consequently, the question of the \emph{optimal} choice of parameters can be posed. Various researchers approach this problem in different ways. In most cases, various linear considerations are employed. For instance, one can try to optimize the linear dispersion relation of the model in regards of the full Euler equations \cite{Madsen2002, Madsen03, Chazel2009}. If the available parameters are several to be optimized, one can employ also the considerations of the linear wave shoaling. In any case, it is the linear part of the model which gets improved. However, later on this \emph{optimized} model will be used to simulate nonlinear waves. So, it is reasonable to ask whether the nonlinear properties of the model at hands benefited by this improvement? In the present study we will try to shed some light on this question by considering a very important class of nonlinear solutions --- the solitary waves \cite{Sandee1991, Malfliet1992a}, which span completely the dynamics in the integrable case \cite{Miura1976}.

The extended versions of the \acs{SGN} equations with a free parameter have already been proposed on flat \cite{Dias2010} and uneven bottoms \cite{Carmo2013}. However, the authors of previous investigations on this topic did not focus on the solitary wave solutions and their confrontation with the full Euler equations. Moreover, in the present study we propose a simple adaptive strategy to find an \emph{optimal value} of the free parameter at every instance of time. Namely, using a simple spectral analysis we determine the dominant wavenumber in the current state of the system. Then, the optimal value is given by satisfying the condition that our approximate model propagates this wavelength with the exact linear celerity (given by the full Euler). This process is repeated at every time step and it costs roughly the computation of one additional FFT\footnote{The Fast Fourier Transform (FFT).} and of two integrals.

The present manuscript is organized as follows. In the following Section~\ref{sec:model} we formulate the extended \acf{eSGN} system. Numerical results on the solitary waves of the \acs{eSGN} equations are presented in Section~\ref{sec:sw}. A novel adaptive strategy for the optimal choice of the free parameter is described and validated in Section~\ref{sec:adapt}. Finally, the main conclusions and perspectives of this study are outlined in Section~\ref{sec:concl}.


\section{Mathematical model}
\label{sec:model}

For two-dimensional surface water waves propagating in shallow water of constant depth, one can approximate the velocity field by
\begin{equation*}
  u(x,y,t)\ \approx\ \bar{u}(x,t), \qquad v(x,y,t)\ \approx\ -\,(y+d)\,\bar{u}_x
\end{equation*}
where $\,d\,$ is the mean water depth and $\,\bar{u}\,$ is the horizontal velocity averaged over the water column --- \ie $\,\bar{u}\equiv h^{-1}\int_{-d}^\eta u\,\ud y$ --- $\,y=\eta\,$ and $\,y=0\,$ being the equations of the free surface and of the still water level, respectively. The horizontal velocity $\,u\,$ is thus uniform along the water column and the vertical velocity $\,v\,$ is chosen so that the fluid incompressibility if fulfilled. \textsc{Serre} (1953) \cite{Serre1953} derived the following approximate system of equations
\begin{align}
h_t\ +\ \partial_x\!\left[\,h\,\bar{u}\,\right]\, &=\ 0, \label{eq:massse} \\
\partial_t\!\left[\,h\,\bar{u}\,\right]\, +\ \partial_x\!\left[\,h\,\bar{u}^2\,
+\,\half\,g\,h^2\,+\,\third\,h^2\,\gamma\,\right]\, &=\ 0, \label{eq:qdmfluxse} 
\end{align}
where 
\begin{equation}\label{eq:defaccver}
\gamma\ =\ h\,(\bar{u}_x^{\,2}-\bar{u}_{xt}-\bar{u}\/\bar{u}_{xx})\ =\ 2\,h\,\bar{u}_x^{\,2}\ -\ h\,\partial_x\!\left[\,\bar{u}_t\,+\,\bar{u}\,\bar{u}_x\,\right],
\end{equation}
is the vertical acceleration of the fluid at the free surface \cite{Clamond2009}. Physically, equations \eqref{eq:massse} and \eqref{eq:qdmfluxse} describe, respectively, the mass and momentum flux conservations. From these two conservative equations, secondary ones can be easily derived using some formal algebraic manipulations:
\begin{align*}
\partial_t\!\left[\,\bar{u}\,-\,\third\,h^{-1}(h^3\/\bar{u}_x)_x\,\right]\, +\ 
\partial_x\!\left[\,\half\,\bar{u}^2\,+\,g\,h\,-\,\half\,h^2\,\bar{u}_x^{\,2}\,-\,
\third\,\bar{u}\,h^{-1}(h^3\/\bar{u}_x)_x\,\right]\, &=\ 0, \\
\partial_t\!\left[\,h\,\bar{u}\,-\,\third(h^3\bar{u}_x)_x\,\right]\, +\ \partial_x\!\left[\,h\,\bar{u}^2\,+\,\half\,g\,h^2\,-\,\twothird\,h^3\,\bar{u}_x^{\,2}\,-\,
\third\,h^3\,\bar{u}\/\bar{u}_{xx}\,-\,h^2\/h_x\/\bar{u}\/\bar{u}_x\,\right]\, &=\ 0, \\ 
\partial_t\!\left[\,\half\,h\,\bar{u}^2\,+\,\sixth\,h^3\/\bar{u}_x^{\,2}\, + \,
\half\,g\,h^2\,\right]\, +\ \partial_x\!\left[\,(\half\,\bar{u}^2\, + \,\sixth\,h^2\, \bar{u}_x^{\,2}+\,g\,h\, + \,\third\,h\,\gamma\,)\,h\,\bar{u}\,\right]\, &=\ 0, 
\end{align*}
We can also rewrite these equations in equivalent non-conservative forms, for instance we have
\begin{equation*}
  \bar{u}_t\ +\ \bar{u}\,\bar{u}_x\ +\ g\,h_x\ +\ \third\,h^{-1}\,\partial_x\!\left[\, h^2\,\gamma\,\right]\, =\ 0,
\end{equation*}
These approximations are valid in shallow water without assuming small amplitude waves, they are therefore sometimes called {\em weakly-dispersive fully-nonlinear approximation} \cite{Wu2001a} and are a generalisation of the Saint--Venant and of the Boussinesq equations.


\subsection{Extended \acl{SGN}'s equation}
\label{sec:ext}

Since the \acs{SGN} equations represent long waves in shallow water, this means that the horizontal and temporal derivatives are small quantities, \ie $\partial_x\propto\mathrm{O}(\eps)$ and $\partial_t\propto \O(\eps)$, where $\eps$ is of the order of the water depth divided by the characteristic wavelength. Introducing explicitly this small parameter, \ie using the scaled variables
\begin{equation*}
  x^\star\ =\ \eps\,x, \qquad t^\star\ =\ \eps\,t, \qquad \gamma^\star\ =\ \eps^{-2}\,\gamma
\end{equation*}
since the vertical acceleration is of order 2 --- \ie, $\gamma\propto\O(\eps^2)$ --- as it is obvious from the definition \eqref{eq:defaccver}, the \acs{SGN} equations can be written as
\begin{align*} 
\epsilon\,h_{t^\star}\ +\ \epsilon\,\partial_{x^\star}\!\left[\,h\,\bar{u}\,\right]\, &=\ 0, 
\\
\epsilon\,\partial_{t^\star}\!\left[\,h\,\bar{u}\,\right]\, +\ \epsilon\,\partial_{x^\star}
\!\left[\,h\,\bar{u}^2\,+\,\half\,g\,h^2\,+\,\epsilon^2\,\third\,h^2\,\gamma^\star\,\right]\, &=\ 0, 
\end{align*}
where all the variables in these equations are of order $\O(1)$. Note that \acs{SGN} equations neglect all terms involving powers of $\epsilon$ higher than three.

Substituting the relation
\begin{equation*}
  \bar{u}_{t^\star}\ +\ \bar{u}\,\bar{u}_{x^\star}\ =\ -\,g\,h_{x^\star}\ -\ \eps^2\,\third\,h^{-1}\,\partial_{x^\star}\!\left[\, h^2\,\gamma^\star\,\right],
\end{equation*}
into the definition of the vertical acceleration $\gamma$, we have
\begin{equation}\label{eq:defaccvermod}
  \gamma\ =\ \eps^2\,2\,h\,\bar{u}_{x^\star}^{\,2}\ +\ \eps^2\,g\,h\,h_{x^\star x^\star}\ +\ \O\!\left(\eps^4\right),
\end{equation}
which is a new expression for the vertical acceleration consistent with the order of approximation.

It is however possible to obtain a more general system averaging the two expressions \eqref{eq:defaccver} and \eqref{eq:defaccvermod}, we have
\begin{equation*}
  \gamma\ =\ \eps^2\,2\,h\,\bar{u}_{x^\star}^{\,2}\ +\ (1-\alpha)\,\eps^2\,g\,h\, h_{x^\star x^\star}\ -\ \alpha\,\eps^2\,h\,\partial_{x^\star}\!\left[\,\bar{u}_{t^\star}\, + \,\bar{u}\,\bar{u}_{x^\star}\,\right]\,+\ \O\!\left(\eps^4\right),
\end{equation*}
where $\alpha$ is a constant at our disposal. Thence, returning to the original variables, the modified \acs{SGN}'s equations are
\begin{align}
  h_t\ +\ \partial_x\!\left[\,h\,\bar{u}\,\right]\, &=\ 0,  \label{eq:masssem} \\
  \partial_t\!\left[\,h\,\bar{u}\,\right]\, +\ \partial_x\!\left[\,h\,\bar{u}^2\, +\,\half\,g\,h^2\,+\,\third\,h^2\,\gamma\,\right]\, &=\ 0, \label{eq:qdmfluxsem} \\
  2\,h\,\bar{u}_x^{\,2}\ +\ (1-\alpha)\,g\,h\,h_{xx}\ -\ \alpha\,\,h\,\partial_x\!\left[\,\bar{u}_t\,+\,\bar{u}\,\bar{u}_x\,\right]\,&=\ \gamma.
\label{eq:veraccmod}
\end{align}
From these modified \acs{SGN}'s equations, we can derive a secondary relation, which can be interpreted physically as the horizontal momentum conservation law:
\begin{align*}
  \partial_t\!\left[\,h\,\bar{u}\,-\,\third\,\alpha\/(h^3\bar{u}_x)_x\,\right]\, +\ \partial_x\!\left[\,h\,\bar{u}^2\,+\,\half\,g\,h^2\,+\,\third\,(1-\alpha)\,g\,h^3\,h_{xx} \quad\right. \\ \left.+\,\twothird\,(1-2\alpha)\,h^3\,\bar{u}_x^{\,2}\,-\,\third\,\alpha\,h^3\,\bar{u}\,\bar{u}_{xx}\,-\,\alpha\,h^2\,h_x\,\bar{u}\,\bar{u}_x\,\right]\, &=\ 0. 
\end{align*}
It can be recast equivalently as a system of two equations, which are more convenient for numerical computations:
\begin{align*}
  q_t\ +\ \partial_x\!\left[\,\bar{u}\,q\,+\,\half\,g\,h^2\,+\,\third\,(1-\alpha)\,g\,h^3\,h_{xx}\, + \,\twothird\,(1-2\alpha)\,h^3\,\bar{u}_x^{\,2}\,\right]\, &=\ 0, \\
  h\,\bar{u}\ -\ \third\,\alpha\,\partial_x\!\left[\,h^3\,\bar{u}_x\,\right]\, &= \ q.
\end{align*}

\begin{remark}
We were not able to find a variational (Lagrangian or Hamiltonian) structure of governing equations \eqref{eq:masssem}, \eqref{eq:veraccmod}. The derivation of extended \acs{SGN} equations possessing such a structure will be one of the challenges we will address in upcoming studies.
\end{remark}


\subsection{Linear approximation}

For infinitesimal waves, $\eta$ and $\bar{u}$ being both small, it is reasonable to linearise the equations around $\eta = 0$ and $\bar{u} = 0$. We obtain thus the linear system of equations
\begin{align}
  \eta_t\ +\ d\,\bar{u}_x\ &=\ 0,  \label{eqmasssemlin} \\
  \bar{u}_t\ +\ g\,\eta_x\ +\ \third\,d\,\gamma_x\ &=\ 0, \label{eqqdmfluxsemlin} \\
  (1-\alpha)\,g\,d\,\eta_{xx}\ -\ \alpha\,\,d\,\bar{u}_{xt}\ &=\ \gamma. \label{eqveraccmod}
\end{align}
Seeking for traveling waves of the form $\eta=a\cos \bigl(k(x-ct)\bigr)$, we obtain the (linear) dispersion relation
\begin{equation}\label{eq:disrelserlin}
  \frac{c^2}{g\/d}\ =\ \frac{3\,+\,(\alpha-1)\/(k\/d)^2}{3\,+\,\alpha\/(k\/d)^2}\ =\ 1\, -\, \third\/(k\/d)^2\, +\,\textstyle{1\over9}\/\alpha\/(k\/d)^4\,-\, \textstyle{1\over27}\/\alpha^2\/(k\/d)^6\,+\,\cdots.
\end{equation}
We note that this relation is well-posed (\ie $c^2>0$ for all $k$) only if $\alpha \geqslant 1$. In order to find a suitable choice for $\alpha$, the relation \eqref{eq:disrelserlin} can be compared with the dispersion relation of linear waves on finite depth
\begin{equation}\label{eq:disrelexa}
  c^2\,/\,g\,d\ =\ \mathrm{thc}(k\/d)\ =\ 1\, -\, \third\/(k\/d)^2\, +\,\textstyle{2\over15}\/(k\/d)^4\,-\,\textstyle{17\over315}\/(k\/d)^6\, + \,\textstyle{62\over2835}\/(k\/d)^8\, + \,\cdots,
\end{equation}
where $\mathrm{thc}(x)\equiv\tanh(x)/x$ if $x \neq 0$ and $\mathrm{thc}(0) \equiv 1$. Comparing the Taylor expansions, it is clear that \eqref{eq:disrelserlin} matches the exact one only up to the second-order in general, except when $\alpha=6/5$ in which case it matches up to the fourth-order. Therefore, $\alpha_{\mathrm{opt}} = 6/5$ is a suitable choice having the advantage of being independent of the wave characteristics. This method of choosing the optimal $\alpha$ has been used by many authors starting from the pioneering works \cite{BS, Nwogu1993, Beji1996}.

Let us discuss some other possible choices of the free parameter $\alpha$. We may choose $\alpha$ such that the dispersion relations \eqref{eq:disrelserlin} and \eqref{eq:disrelexa} are equal, \ie, such that
\begin{equation}\label{eq:eqal}\tag{$\star$}
  \frac{3\,+\,(\alpha-1)\/(k\/d)^2}{3\,+\,\alpha\/(k\/d)^2}\ =\ \frac{\tanh(k\/d)}{k\/d},
\end{equation}
thence
\begin{equation}\label{eq:expandalpha}
  \alpha\ =\ \frac{(kd)^2\,-\,3\,(1-\mathrm{thc}(kd))}{(kd)^2\,(1-\mathrm{thc}(kd))}\ =\ \frac{6}{5}\ -\ \frac{(kd)^2}{175}\ +\ \frac{2\,(kd)^4}{7875}\ -\ \cdots.
\end{equation}
This choice of $\alpha$ is suitable for periodic waves when the wavelength $kd$ is given. When the celerity is given, it is more suitable to proceed as follows. Solving \eqref{eq:disrelserlin} for $k$, one gets
\begin{equation*}
  (kd)^2\ =\ \frac{3\,(c^2-gd)}{(\alpha-1)\,g\,d\,-\,\alpha\,c^2},
\end{equation*}
and reporting into \eqref{eq:disrelexa}, one obtains a transcendent equation for $\alpha$ which has to be solved numerically using some fixed point or Newton-type iterations \cite{Isaacson1966}:
\begin{equation*}
  \frac{c^2}{g\/d}\ =\ \mathrm{thc}\!\left(\sqrt{\frac{3\,(c^2-gd)}{(\alpha-1)\,g\,d\,-\,\alpha\,c^2}}\,\right).
\end{equation*}

\begin{remark}
Consider now a steady wave motion, \ie, solution independent of time. We can easily derive the formulation for steady waves as well from the governing equations. The mass conservation \eqref{eq:masssem} yields
\begin{equation*}
  \bar{u}\ =\ -\,c\,d\left/\,h\right.,
\end{equation*}
and substituting into \eqref{eq:qdmfluxsem} and \eqref{eq:veraccmod}
\begin{equation}\label{eq:bab}
  \frac{c^2}{g\,h}\ +\ \frac{h^2}{2\,d^2}\ +\ \frac{\gamma\,h^2}{3\,g\,d^2}\ =\ \frac{c^2}{g\,d}\ +\ \frac{1}{2}\ +\ K
\end{equation}
where $K$ is a (dimensionless) integration constant ($K = 0$ for solitary waves).
\end{remark}

\begin{remark}\label{rem:sw}
Solitary waves for the classical \acs{SGN} equations are known analytically:
\begin{equation}\label{eq:swsol}
  \eta = a\ \sech^2\half \kappa(x-ct), \quad
  \bar{u} = \frac{c\,\eta}{d+\eta}, \quad 
  c^2 = g(d+a), \quad 
  (\kappa d)^2 = \frac{3a}{d+a}.
\end{equation}
Unfortunately, for a generic value of $\alpha$ there are no exact solutions known for the \acs{eSGN} equations. Consequently, we will employ numerical methods in Section~\ref{sec:sw} to find the travelling waves to high accuracy.
\end{remark}


\section{Numerical methods and results}
\label{sec:num}

In order to study some properties and the performance of the proposed \acs{eSGN} equations we employ numerical methods. We do not enter into the details of the numerical methods here, since they can be found in the literature. For the computation of travelling waves we employ the Levenberg-Marquardt algorithm \cite{More1978}. The main ideas of this method are summarized briefly below. For the transient simulations of the classical \acs{SGN} and new \acs{eSGN} equations we use a pseudo-spectral scheme described in \cite{Dutykh2011a}. For the validations of \acs{eSGN} model predictions, we perform the comparisons with the full Euler equations which are solved using the dynamic conformal mapping technique proposed by L.V.~\textsc{Ovsyannikov} \cite{Ovsyannikov1974} and developed later by several authors \cite{Choi1999, Li2004}.


\subsection{Solitary waves}
\label{sec:sw}

In this Section we will investigate the influence of the parameter $\alpha$ on solitary waves, as the most important class of nonlinear solutions. We recall that the optimal choice of $\alpha$ is directed by some linear considerations and its impact on the nonlinear properties of the \acs{eSGN} system is not obvious.

We will compare the solitary wave solutions to the three following models:
\begin{itemize}
  \item \acs{SGN} equations
  \item \acs{eSGN} equations (with optimal $\alpha$)
  \item the full Euler equations (the reference solution)
\end{itemize}
The solitary waves to the classical \acs{SGN} equations are known analytically (see Remark~\ref{rem:sw}). The solitary wave solutions for the full Euler equations are computed using the method of conformal variables \cite{Clamond2012b, Dutykh2013b}. The \textsc{Matlab} script used to generate the solitary waves can be downloaded at \cite{Clamond2012}. Unfortunately, we did not succeed in finding analytical solutions to the \acs{eSGN} equations for a general $\alpha$. Consequently, we had to employ the numerical methods. 

Equation~\eqref{eq:bab} is discretized in space using the classical Fourier-type pseudo-spectral method \cite{Boyd2000}. For steady computations we did not even find the necessity to employ any anti-aliasing rule. The discrete system was solved using the so-called Levenberg--Marquardt algorithm proposed independently by \textsc{Levenberg} (1944) \cite{Levenberg1944} and \textsc{Marquardt} (1963) \cite{Marquardt1963} who gave the name to this method successfully applied nowadays to various problems \cite{Lourakis2005}. The main idea behind this method is, first, to reformulate the system of equations as a nonlinear least-squares problem. Then, the nonlinear least-squares problem is solved iteratively with the steepest descent method far from the solution and, with the Newton's method in the vicinity of the root, where the convergence will be quadratic. The Jacobian matrix is computed using central finite differences. The initial guess was given by the analytical solution \eqref{eq:swsol}. Only a relatively small number of iterations needed to achieve the convergence (typically less then 20). The computational domain consists of the periodic interval $[-\ell, \ell] = [-30, 30]$ which was discretized using $N = 2048$ equally spaced collocation points.

\begin{figure}
  \centering
  \includegraphics[width=0.79\textwidth]{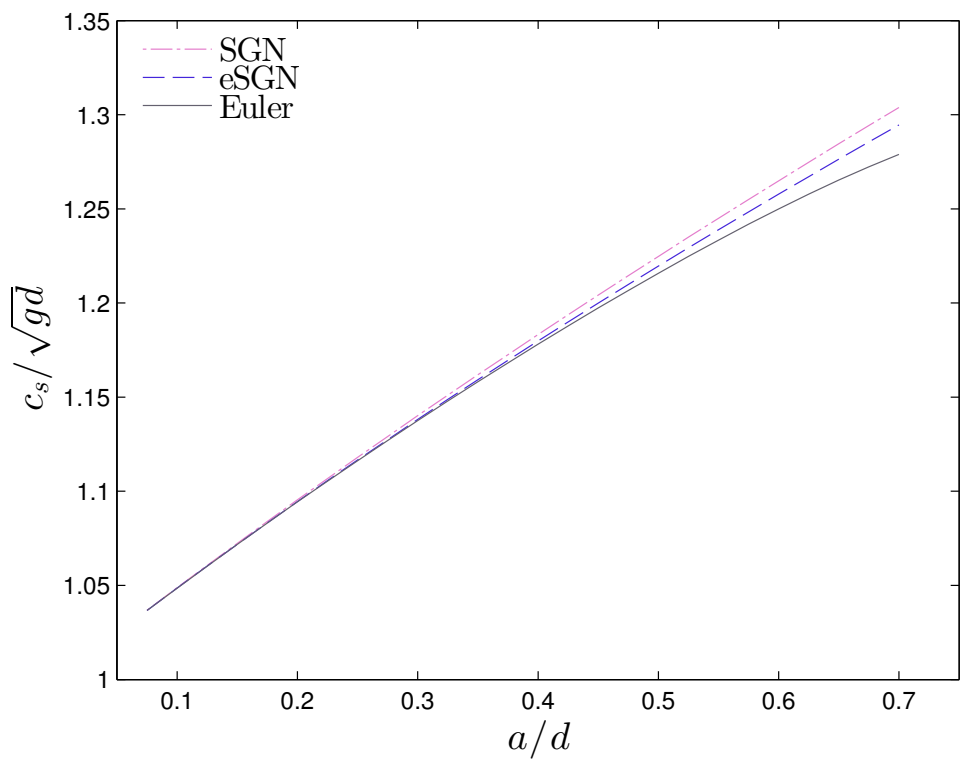}
  \caption{\small\em Speed--amplitude relations for solitary waves in \acs{SGN}, \acs{eSGN} and the full Euler equations ($\alpha = 6/5$).}
  \label{fig:spa}
\end{figure}

\begin{figure}
  \centering
  \subfigure[]{\includegraphics[width=0.48\textwidth]{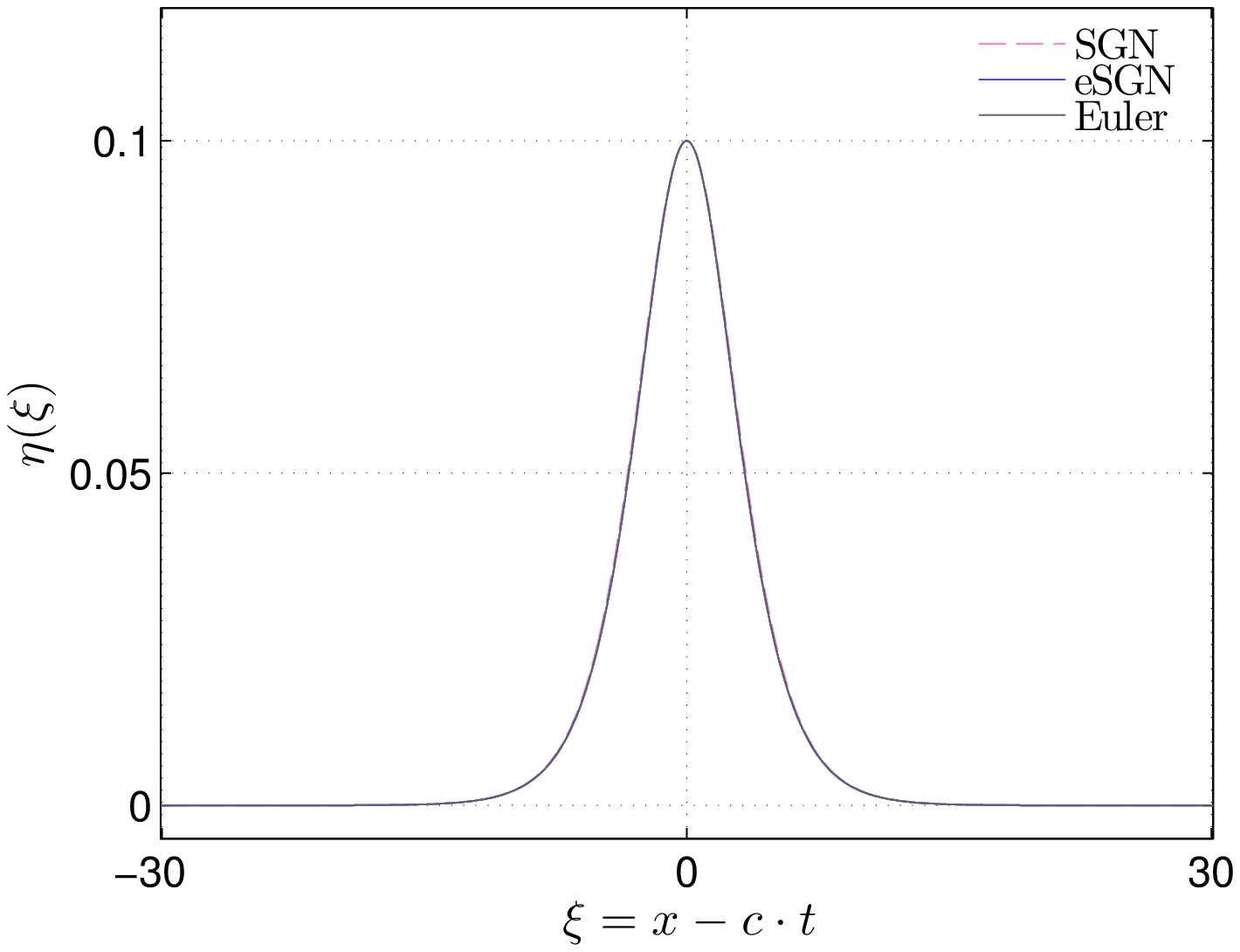}}
  \subfigure[]{\includegraphics[width=0.48\textwidth]{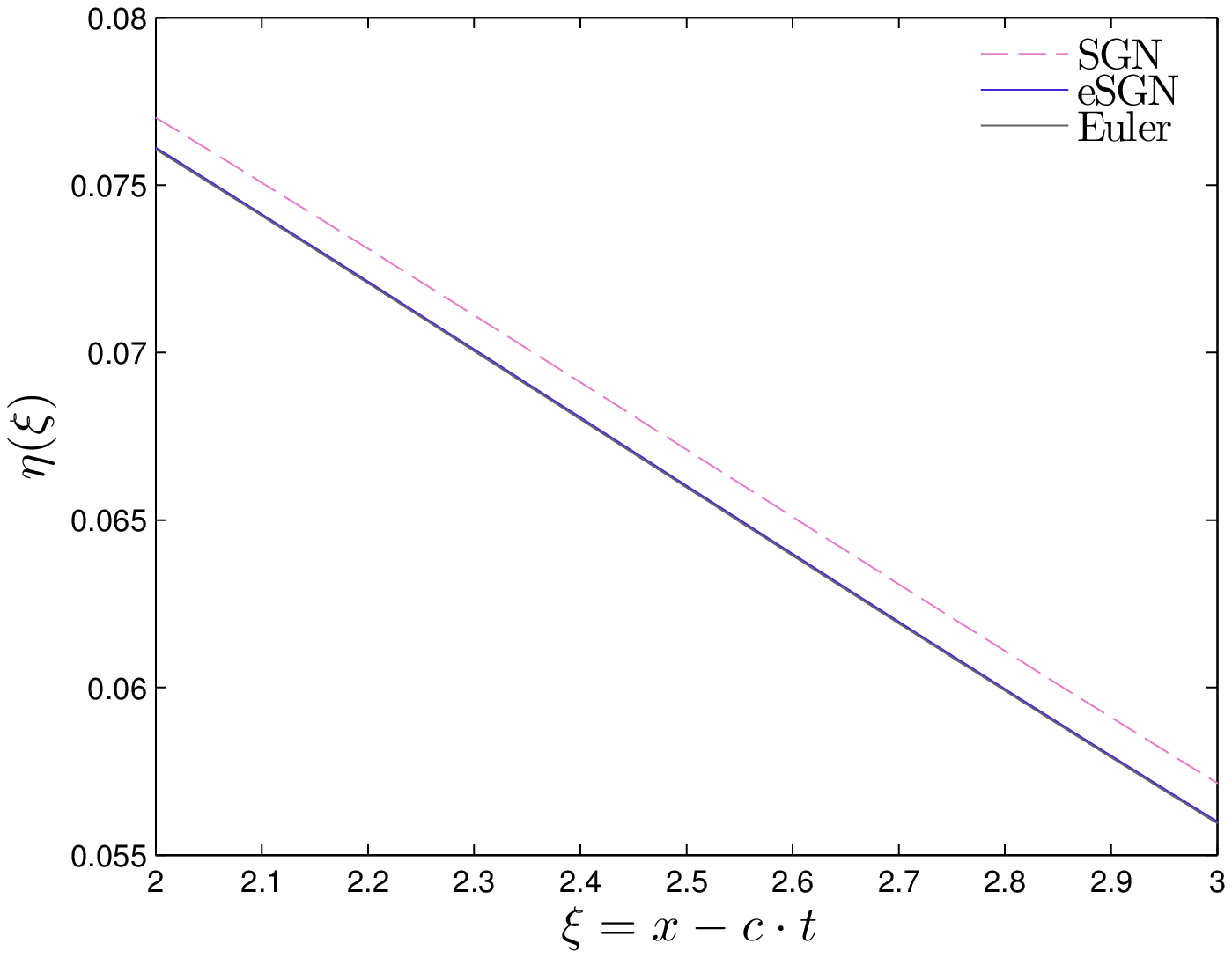}}
  \caption{\small\em Small amplitude solitary wave solutions to the \acs{SGN}, \acs{eSGN} and the full Euler equations of amplitude $a/d = 0.1$. The right panel shows a zoom on $2\leq \xi \leq 3$ ($\alpha = 6/5$).}
  \label{fig:small}
\end{figure}

\begin{figure}
  \centering
  \subfigure[]{\includegraphics[width=0.48\textwidth]{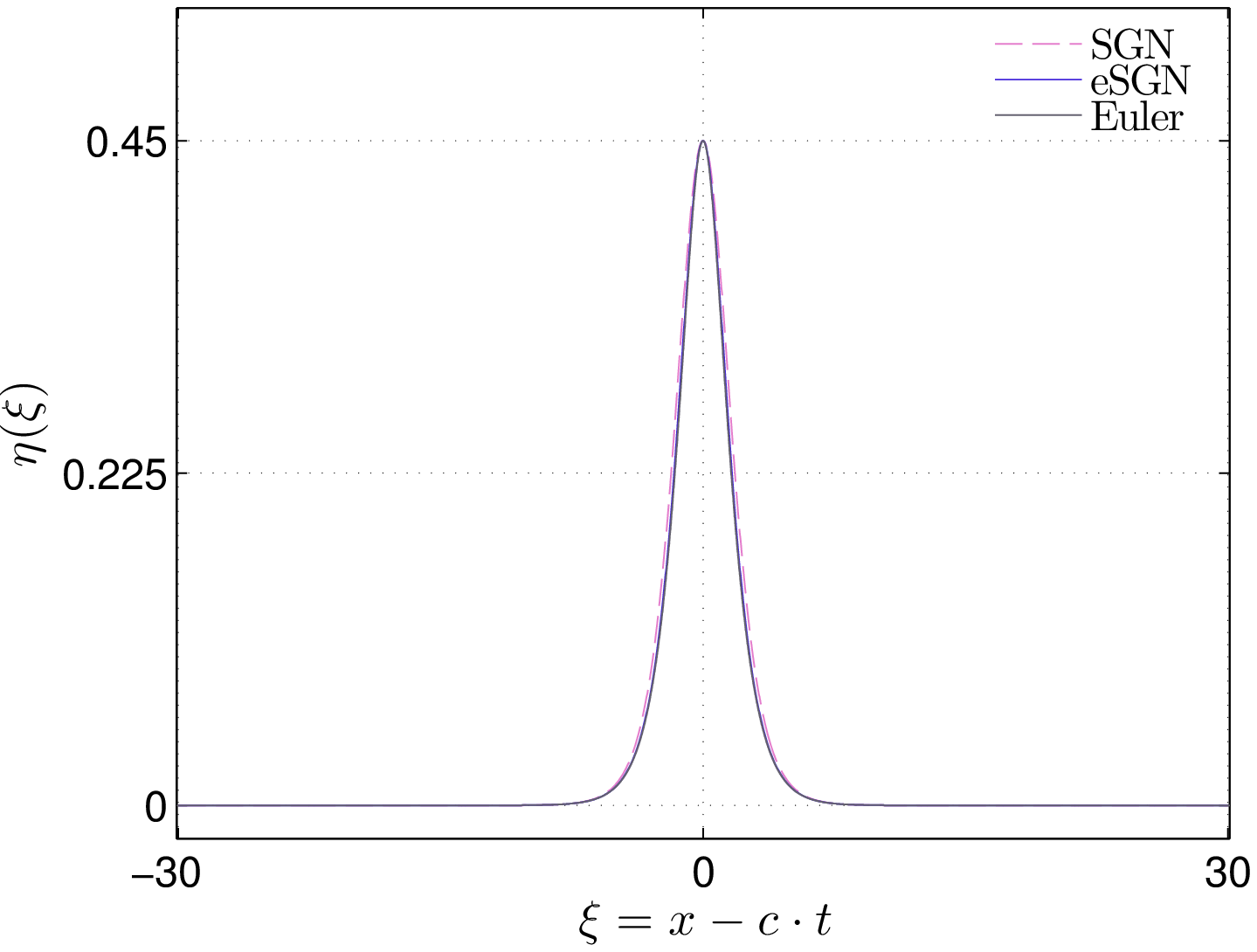}}
  \subfigure[]{\includegraphics[width=0.48\textwidth]{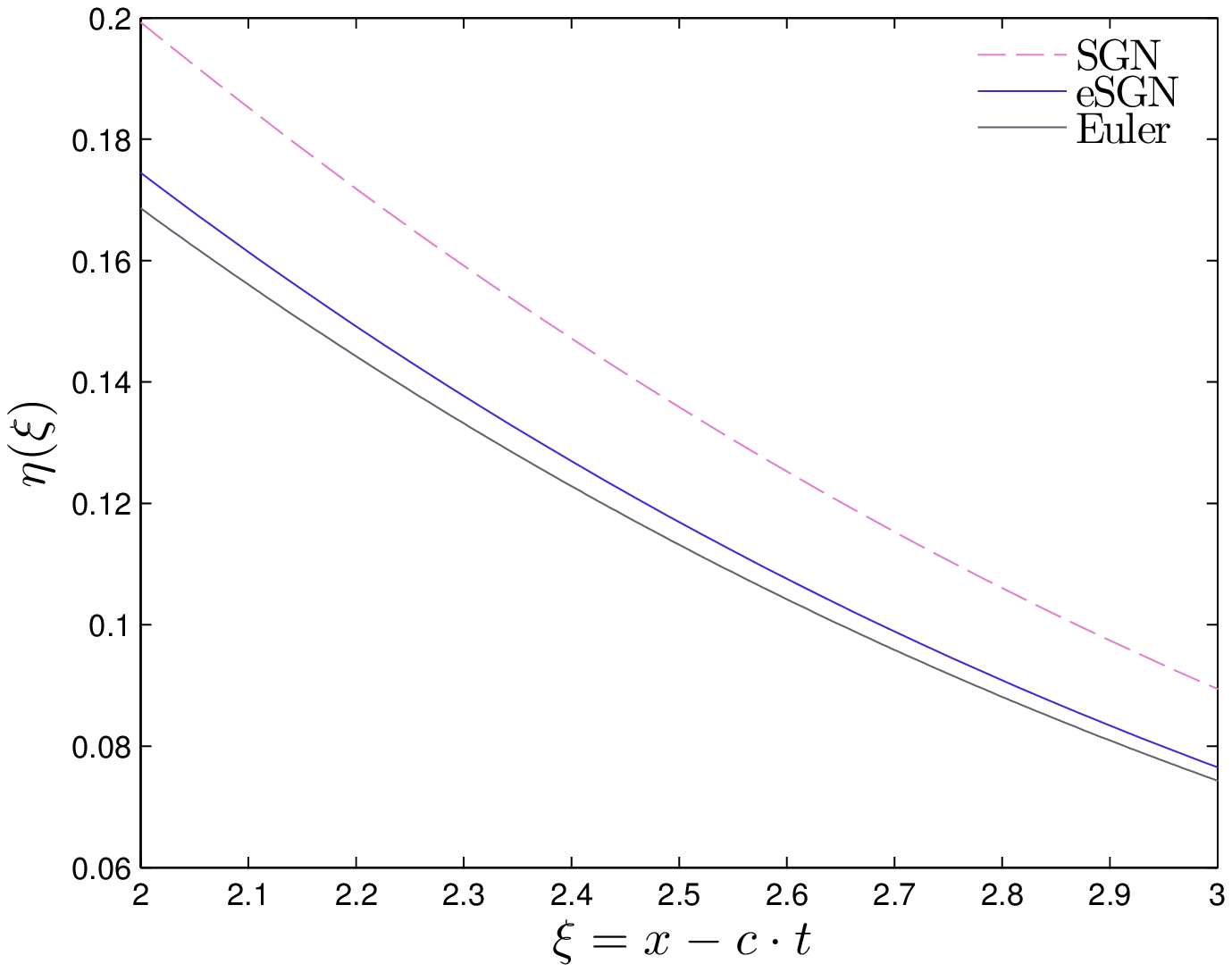}}
  \caption{\small\em Moderate amplitude solitary wave solutions to the \acs{SGN}, \acs{eSGN} and the full Euler equations of amplitude $a/d = 0.45$. The right panel shows a zoom on $2\leq \xi \leq 3$ ($\alpha = 6/5$).}
  \label{fig:medium}
\end{figure}

\begin{figure}
  \centering
  \subfigure[]{\includegraphics[width=0.48\textwidth]{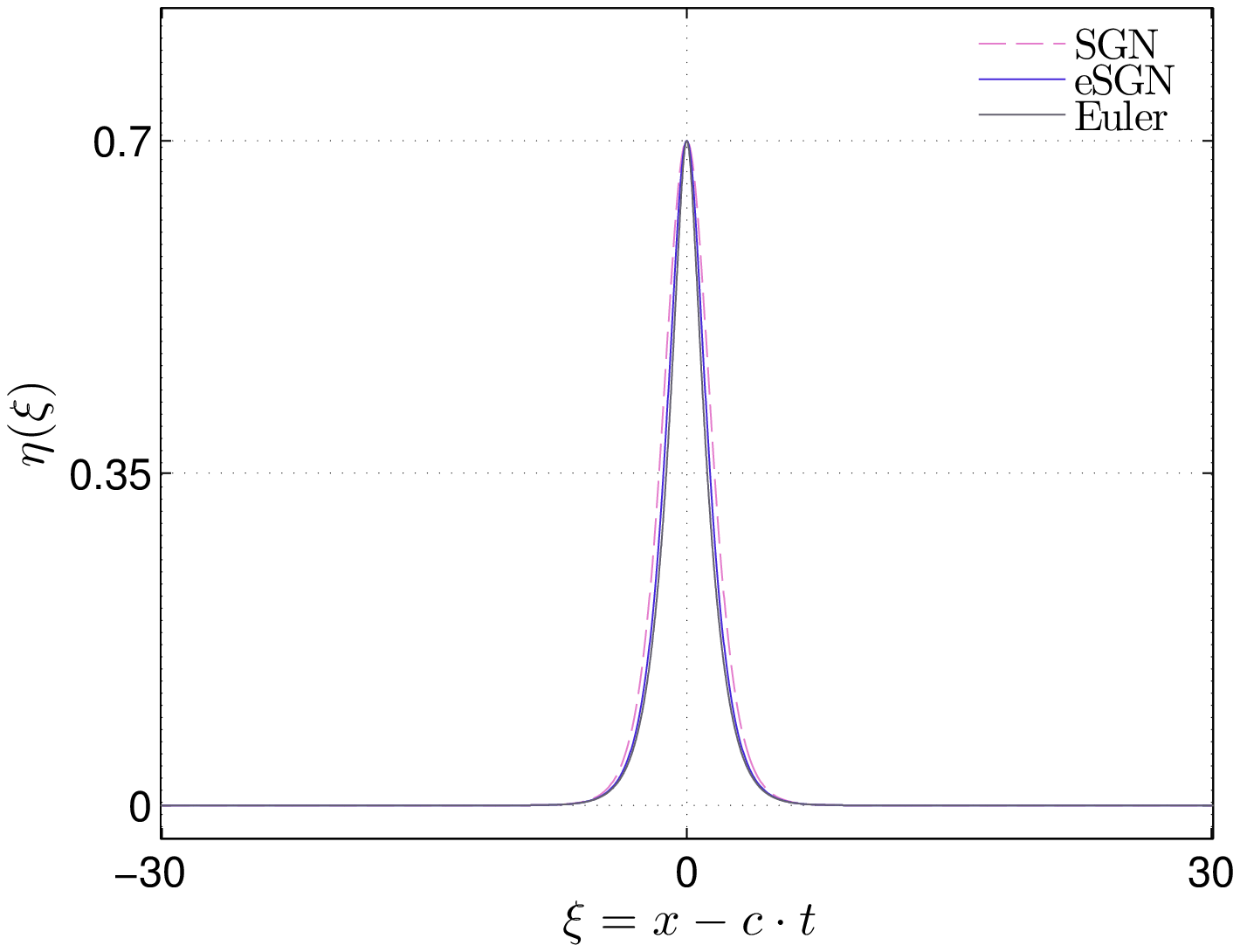}}
  \subfigure[]{\includegraphics[width=0.48\textwidth]{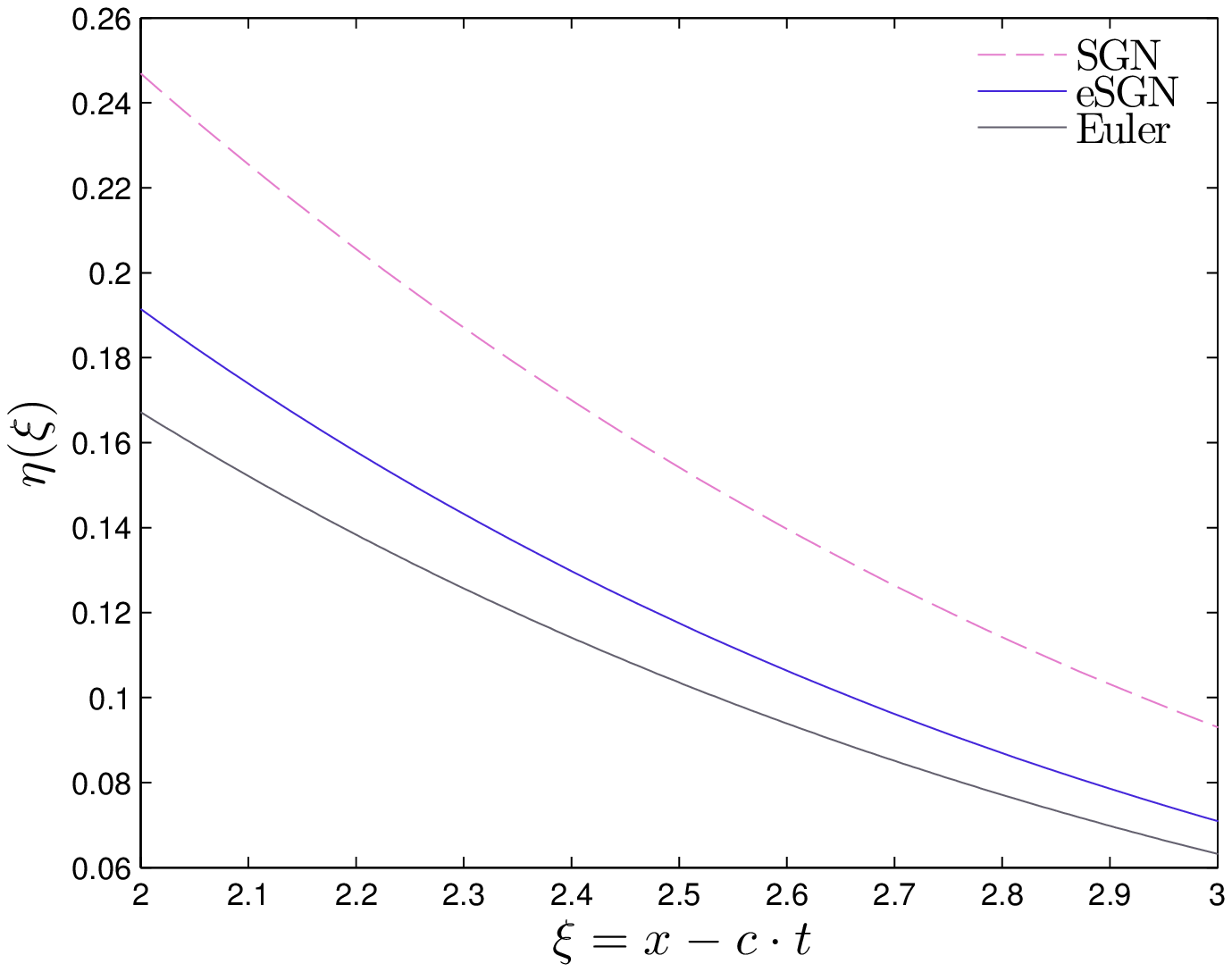}}
  \caption{\small\em Highly nonlinear solitary wave solutions to the \acs{SGN}, \acs{eSGN} and the full Euler equations of amplitude $a/d = 0.7$. The right panel shows a zoom on $2\leq \xi \leq 3$ ($\alpha = 6/5$).}
  \label{fig:high}
\end{figure}

We will consider three amplitudes of solitary waves $a/d = 0.1$ (small), $0.45$ (medium) and $0.7$ (high amplitude). The propagation speeds predicted by various models are reported in Table~\ref{tab:cs}. One can already see that the \acs{eSGN} predictions are always closer to the full Euler equations. We computed the speed--amplitude relation for the whole range of amplitudes (see Figure~\ref{fig:spa}). The numerical results confirm our preliminary conclusions. The shapes of three solitary waves under consideration are presented on Figures~\ref{fig:small} -- \ref{fig:high} respectively. On the left panels (\textit{a}) the whole computational domain is shown and the waves are undistinguishable to the graphical resolution. Consequently, on the right pictures (\textit{b}) we show a magnification of the subdomain $[2, 3]$. At this stage one can see that the \acs{eSGN} model approximates better the reference solution again. For the small amplitude solitary wave ($a/d = 0.1$) the \acs{eSGN} solution is undistinguishable from the Euler solitary wave even on the magnified Figure~\ref{fig:small}(\textit{b}).

We note that similar comparisons between the full Euler and the classical \acs{SGN} equations have been performed also in \cite{Li2004}. However, we focus here on the performance of the extended \acs{SGN} model with respect to its classical counterpart.

\begin{table}
  \centering
  \begin{tabular}{||>{\columncolor[gray]{0.85}}l||>{\columncolor[gray]{0.85}}c||>{\columncolor[gray]{0.85}}c||>{\columncolor[gray]{0.85}}c||}
  \hline\hline
  \textit{Model/Amplitude} & $a/d = 0.1$ & $a/d = 0.45$ & $a/d = 0.70$ \\
  \hline\hline
  \acs{SGN} speed,  $c_s/\sqrt{gd}$ & \textbf{1.048}80  & \textbf{1.2}041 & \textbf{1.}3038 \\ 
  \acs{eSGN} speed, $c_s/\sqrt{gd}$ & \textbf{1.0485}6  & \textbf{1.19}99 & \textbf{1.2}946 \\ 
  Full Euler speed, $c_s/\sqrt{gd}$ & \textbf{1.048548} & \textbf{1.1973} & \textbf{1.2788} \\ 
  \hline\hline
  \end{tabular}
  \bigskip
  \caption{\small\em Comparison of the solitary wave speeds for several fixed values of the wave amplitude. The parameter $\alpha = 6/5$.}
  \label{tab:cs}
\end{table}

\subsection{Adaptive strategy}
\label{sec:adapt}

Now we discuss the choice of the optimal value of the free parameter $\alpha$ for transient wave computations. Assume that at time $t$ we know the free surface elevation $\eta(x,t)$ profile. Then, we compute its Fourier transform in space
\begin{equation*}
  \hat{\eta}(k,t)\ :=\ \F\{\eta(x,t)\}\ =\ \int_\R\, \eta(x,t)\,\ue^{\ui k x}\,\ud x.
\end{equation*}
Then, we can easily compute the power spectrum as well:
\begin{equation*}
  \hat{\S}(k,t)\ :=\ |\hat{\eta}(k,t)|^2.
\end{equation*}
Now we have all the ingredients to estimate the \emph{dominant wavenumber} $k_0$ \cite{Boccotti2000}:
\begin{equation*}
  k_0(t)\ =\ \frac{\int\,k\hat{\S}(k,t)\,\ud k}{\int\,\hat{\S}(k,t)\,\ud k}.
\end{equation*}
The optimal value of $\alpha$ can be obtained by requiring that the \acs{eSGN} system propagates \emph{exactly} the main wavelength corresponding to $k_0(t)$. Mathematically this step is done by solving equation\footnote{Alternatively, the expansion \eqref{eq:expandalpha} can be used to find an approximation to the optimal $\alpha$.} \eqref{eq:eqal} with respect to $\alpha$:
\begin{equation*}
  \frac{3\,+\,(\alpha-1)\cdot\bigl(k_0(t)\/d\bigr)^2}{3\,+\,\alpha\cdot\bigl(k_0(t)\/d\bigr)^2}\ =\ \frac{\tanh\bigl(k_0(t)\/d\bigr)}{k_0(t)\/d}.
\end{equation*}
Since $k_0(t)$ depends on time, so does the optimal value of the free parameter $\alpha(t)$. Then, the \acs{eSGN} model is solved for one time step with the local (in time) estimated optimal value of $\alpha$.

\begin{remark}
We have to mention that when the system contains longer and longer waves, the adaptive strategy will provide us the optimal values of $\alpha$ which will be close to $\alpha_{\mathrm{opt}} = 6/5$ given by the Taylor expansion \eqref{eq:disrelserlin}.
\end{remark}

\begin{figure}
  \centering
  \includegraphics[width=0.75\textwidth]{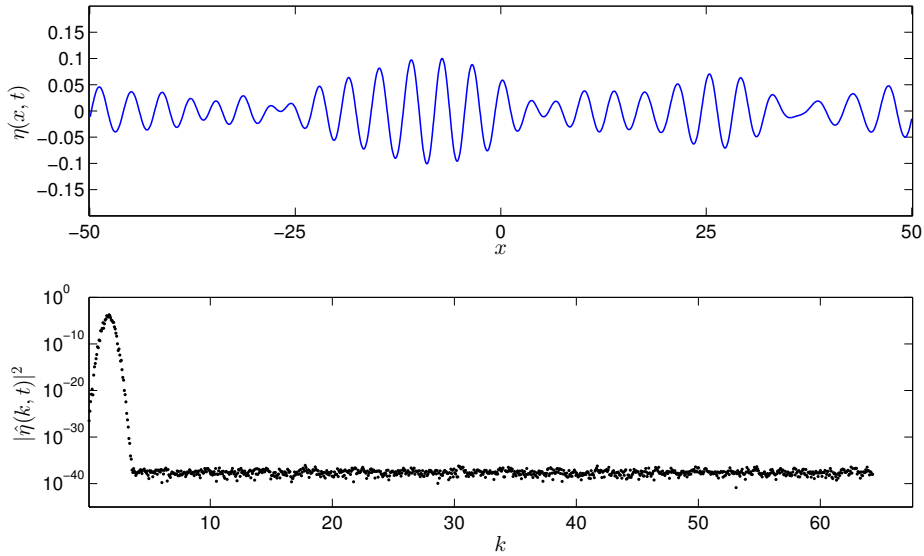}
  \caption{\small\em Gaussian random initial condition used in comparisons. The bottom panel shows the power spectrum $\hat{\S}(k, t = 0)$ of the initial condition. The maximal wave amplitude is $a_0/d = 0.1$.}
  \label{fig:init}
\end{figure}

\begin{figure}
  \centering
  \subfigure[$t = 10.0\ \mathsf{s}$]%
  {\includegraphics[width=0.58\textwidth]{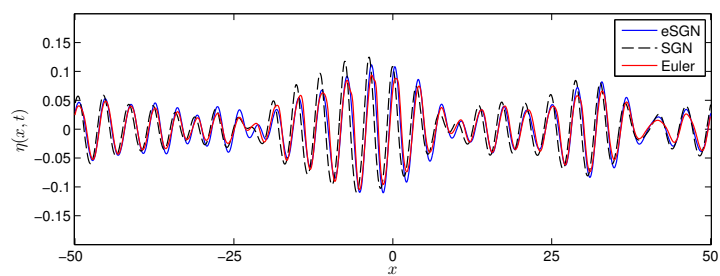}}
  \subfigure[$t = 20.0\ \mathsf{s}$]%
  {\includegraphics[width=0.58\textwidth]{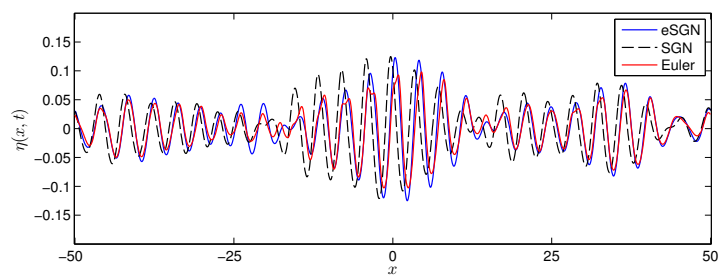}}
  \subfigure[$t = 40.0\ \mathsf{s}$]%
  {\includegraphics[width=0.58\textwidth]{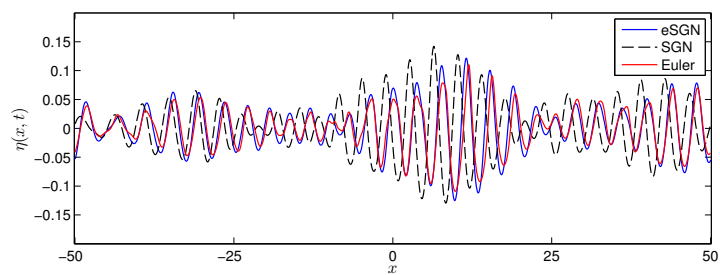}}
  \subfigure[$t = 50.0\ \mathsf{s}$]%
  {\includegraphics[width=0.58\textwidth]{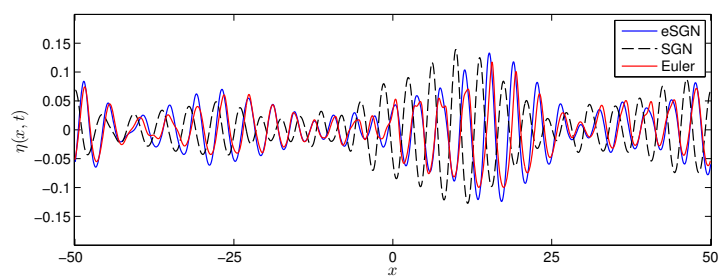}}
  \subfigure[$t = 60.0\ \mathsf{s}$]%
  {\includegraphics[width=0.58\textwidth]{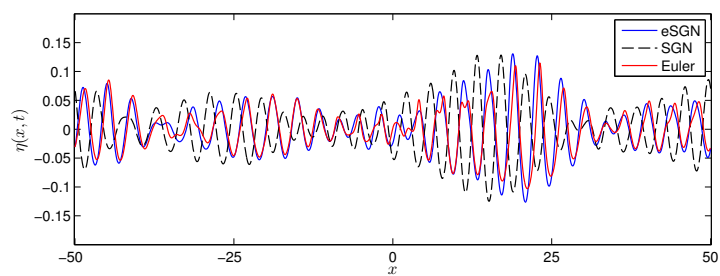}}
  \caption{\small\em Evolution of the initial condition shown on Figure~\ref{fig:init} under the \acs{SGN} (black dashed line), \acs{eSGN} (blue solid line) and the full Euler (red solid line).}
  \label{fig:dyn}
\end{figure}

\begin{figure}
  \centering
  \includegraphics[width=0.99\textwidth]{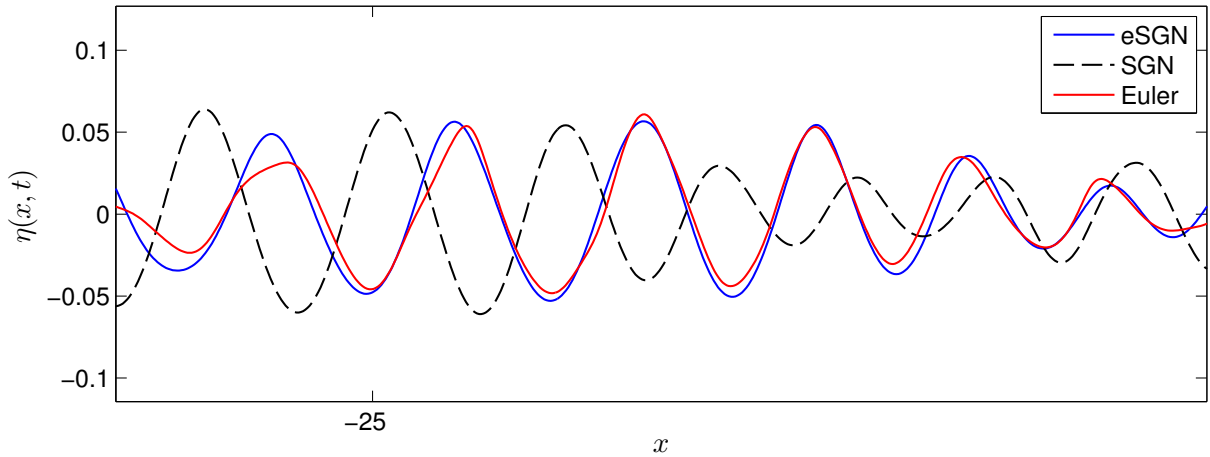}
  \caption{\small\em A zoom on the free surface elevation at the final simulation time $t = T$. Comparison among three models: \acs{SGN} (black dashed line), \acs{eSGN} (blue solid line) and the full Euler (red solid line).}
  \label{fig:zoom}
\end{figure}

\begin{figure}
  \centering
  \includegraphics[width=0.75\textwidth]{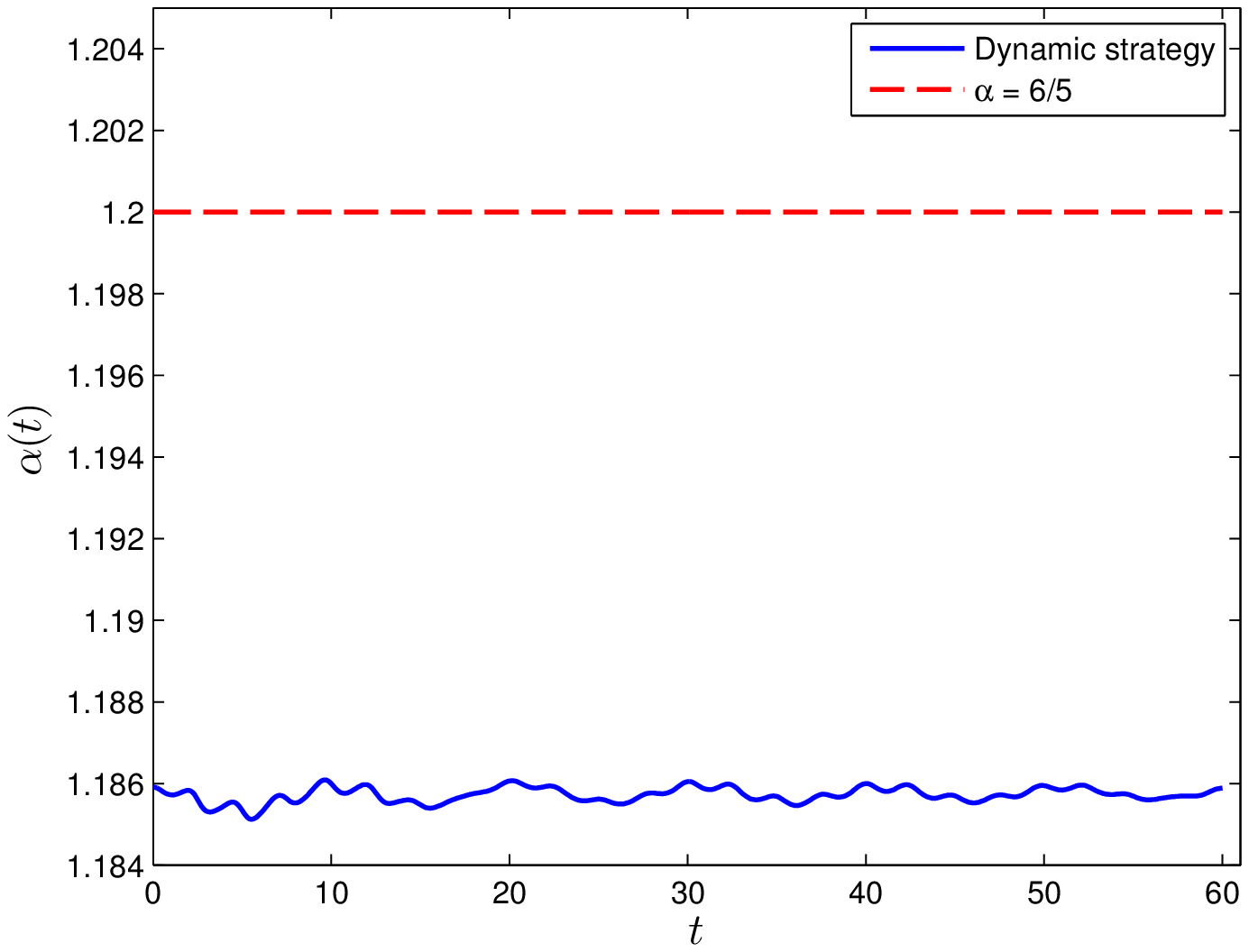}
  \caption{\small\em Evolution of the optimal value of the free parameter $\alpha(t)$ as a function of time (blue solid line). The red dotted line shows the optimal value $\alpha_{\mathrm{opt}} = 6/5$ given by identifying the coefficients of Taylor expansions of the phase velocity. The mean value of $\langle\alpha(t)\rangle \approx 1.1857$ on this trajectory.}
  \label{fig:alpha}
\end{figure}

In order to test the proposed methodology, we perform the comparisons among the three models already studied in Section~\ref{sec:sw} in the steady case. However, this time we consider dynamic (transient) solutions. For this purpose we generate a random Gaussian sea state with the mean wavelength $\lambda_0$ and variance $\sigma_0$. The phases are uniformly distributed random numbers in $[0, 2\pi)$. The values of all physical and numerical parameters are given in Table~\ref{tab:params}. The initial random condition used in our simulations is shown on Figure~\ref{fig:init}. The initial nonlinearity parameter is $\eps = a_0/d = 0.1$ and the shallowness is $\mu^2 = \bigl(\frac{d}{\lambda_0}\bigr)^2 = 0.0625$. The evolution of this initial condition on time horizon $[0, Tf]$ is shown on Figure~\ref{fig:dyn}. We can see that both \acs{SGN} and \acs{eSGN} models do not represent correctly the wave amplitudes and the asymmetry of waves. However, directly from the beginning ($t = 10.0$) the \acs{SGN} solution starts lagging behind the Euler's solution. When the time evolves, this difference accumulates and becomes clearly visible (\eg see Figure~\ref{fig:dyn}(\textit{e})). The \acs{eSGN} solution follows the full Euler much closer. In order to appreciate better the model performance of the proposed adaptive strategy we show also a magnification of the free surface elevation at the final time $t = T$ on Figure~\ref{fig:zoom}. This success is explained by the appropriate and judicious choice of the free parameter $\alpha(t)$. The evolution of $\alpha(t)$ in course of the simulation is shown in Figure~\ref{fig:alpha}. The mean value of the parameter is $\langle\alpha(t)\rangle \approx 1.1857$ on this trajectory. The difference with $\alpha_{\mathrm{opt}} = 6/5 = 1.2$ is not enormous, however it is crucial to represent correctly the wave front positon. A similar animation for a different initial condition can be watched at the following URL address:
\begin{center}
  \url{http://youtu.be/NfgLs7c1keU/}
\end{center}

\begin{table}
  \centering
  \begin{tabular}{||>{\columncolor[gray]{0.85}}c||>{\columncolor[gray]{0.85}}c||}
  \hline\hline
  \textit{Parameter} & \textit{Value} \\
  \hline\hline
  Undisturbed water depth, $d$ $\mathsf{m}$ & 1.0 \\
  Gravity acceleration, $g$ $\mathsf{m}/\mathsf{s}^2$ & 1.0 \\
  Mean wavelength, $\lambda_0$ $\mathsf{m}$ & 4.0 \\
  Initial dominant wavenumber, $k_0 = \frac{2\pi}{\lambda_0}$ $\mathsf{m}^{-1}$ & 1.57 \\
  Variance of the wave spectrum, $\sigma_0$ & 0.1 \\
  Nonlinearity parameter, $\eps = \frac{a_0}{d}$ & 0.1 \\
  Shallowness parameter, $\mu^2 = \bigl(\frac{d}{\lambda_0}\bigr)^2$ & 0.0625 \\
  Computational domain half-length, $\ell$ $\mathsf{m}$ & 50.0 \\
  Number of Fourier modes, $N$ & 2048 \\
  Final simulation time, $T$ $\mathsf{s}$ & 60.0 \\
  \hline\hline
  \end{tabular}
  \bigskip
  \caption{\small\em Physical and numerical parameters used for random wave evolution simulations.}
  \label{tab:params}
\end{table}

\section{Discussion}
\label{sec:concl}

Below we will briefly outline the main conlcusions and perspectives which are opened after the present study.

\subsection{Conclusions}

In this paper we discussed a particular extension of the \acf{SGN} system which is based on the Bona--Smith--Nwogu trick \cite{BS, Nwogu1993}. This idea is not new and the main contribution of this study is not there. Once the free parameter was introduced into the model, we have to provide some recommendations for the practitioners on the choice of this parameter. Most of works available in the literture involve some linear considerations, such as the widely used Taylor expansion or some other optimisation-based procedures of the linear dispersion relation \cite{Madsen2002, Madsen03, Chazel2009}. It is reasonable to question the influence of this \emph{optimal} choice on the nonlinear properties of the model, since it is used to simulate nonlinear waves. In this manuscript we showed that the optimal value of $\alpha$ obtained by the Taylor expansion method leads also to a serious improvement in the solitary wave solutions as well. They approximate much better the corresponding solutions to the full Euler equations comparing to the original \acs{SGN} equations. Namely, the shape as well as the speed--amplitude relation are greatly improved, especially for high amplitude waves. The price to pay for this improvement is that the order of spatial derivatives in the model is increased by one. So, it is up to every user to decide whether this price is worth paying it for the extra accuracy.

The Taylor expansion \eqref{eq:disrelserlin} is valid strictly speaking only in the vicinity of $kd = 0$, \ie infinitely long waves. Unfortunately, such waves cannot be encountered in practice, since every wave has its well-defined wavelength. Consequently, for practical simulations the scheme described above had to be modified to integrate the knowledge of the finite wavelength. In this way we introduced an adaptive strategy which estimates on every time step the dominant wavenumber (\ie the wavelength) and adapts the system to be as accurate as possible for the main spectral component (which carries most of the energy). Our comparisons with the full Euler equations show that this strategy leads to a significant improvement of the \acs{eSGN} model accuracy compared to the classical \acs{SGN} equations. To our knowledge, it is the first study where such an adaptive strategy is proposed and validated.

\subsection{Perspectives}

The present article is only the first step towards the development of the physically adaptive water wave modeling. Further validations are needed, even if the preliminary results are very promising. As the next step, the \acs{eSGN} model has to be generalized to uneven bottoms. However, there are more serious issues with the proposed strategy. The dynamic adaptation introduces time-dependent coefficient $\alpha(t)$ into the PDEs. It implies that we break the invariance of the governing equations with respect to time translations. By Noether theorem\footnote{Rigorously speaking, in order to be able to apply the Noether theorem, the governing equations need to have the Lagrangian structure. However, the absence of the invariance with respect to translations in time allows to conclude.}, we cannot expect the \acs{eSGN} system to conserve exactly the energy. This issue is to be addressed in future investigations.

We could see that in our conservative simulations the dominant wave number did not vary too much on the time horizon considered in the present study. Consequently, we could replace $\alpha(t) \approx \alpha(0)$ determined from the initial condition. However, in the presence of (wind) forcing and/or viscous dissipation, reflective boundary conditions could lead to more drastic modifications of the wave spectrum on longer time scales. Consequently, ``\emph{freezing}'' of the free parameter $\alpha(t)$ cannot be seen as a universal solution.

Another possible drawback comes from the fact that the \acs{eSGN} system \eqref{eq:masssem}, \eqref{eq:qdmfluxsem} was derived under an implicit assumption that the parameter $\alpha$ is constant. Then we allow this parameter to vary with in time. Perhaps, our system discards some terms proportional to $\dot{\alpha}(t)$. However, our numerical results show (see Figure~\ref{fig:alpha}) that $\alpha(t)$ does not vary significantly in time. Hence, the discarded terms can be effectively neglected $\dot{\alpha}(t) \approx 0$. However, we would like to clarify completely this situation in future studies.

Nevertheless, the gain in accuracy we witnessed in the \acs{eSGN} system when it is supplemented with the adaptive strategy, certainly overbalance the shortcomings mentioned hereinabove.


\subsection*{Acknowledgments}
\addcontentsline{toc}{subsection}{Acknowledgments}

The authors would like to thank Professors Angel~\textsc{Duran} (University of Valladolid, Spain) and Vadym~\textsc{Aizinger} (Friedrich-Alexander Universit\"at Erlangen-N\"urnberg, Germany) for very stimulating discussions on the numerical methods for nonlinear waves and adaptive physical models. D.~\textsc{Mitsotakis} was supported by the Marsden Fund administered by the Royal Society of New Zealand.


\addcontentsline{toc}{section}{References}
\bibliographystyle{abbrv}
\bibliography{biblio}

\end{document}